\documentclass[aps,prb,twocolumn,superscriptaddress,floatfix,amsmath,amssymb]{revtex4-2}
\usepackage{amsmath}
\usepackage{amssymb}
\usepackage{graphicx}
\usepackage{float}

\begin{document}
\title{Band alignment and interlayer hybridisation in transition metal dichalcogenide/hexagonal boron nitride heterostructures}

\author{S. J. Magorrian}
\email{samuel.magorrian@warwick.ac.uk}
\affiliation{Department of Physics, University of Warwick, Coventry, CV4 7AL, United Kingdom}
\author{A. J. Graham}
\affiliation{Department of Physics, University of Warwick, Coventry, CV4 7AL, United Kingdom}
\author{N. Yeung}
\affiliation{Department of Physics, University of Warwick, Coventry, CV4 7AL, United Kingdom}
\author{F. Ferreira}
\affiliation{National Graphene Institute, University of Manchester, Booth St E, Manchester, M13 9PL, United Kingdom}
\affiliation{Department of Physics \& Astronomy, University of Manchester, Oxford Road, Manchester, M13 9PL, United Kingdom}
\author{P. V. Nguyen}
\affiliation{Department of Physics, University of Washington, Seattle, WA, USA}
\author{A. Barinov}
\affiliation{Elettra—Sincrotrone Trieste, S.C.p.A., Basovizza (TS) 34149, Italy}
\author{V. I. Fal'ko}
\affiliation{National Graphene Institute, University of Manchester, Booth St E, Manchester, M13 9PL, United Kingdom}
\affiliation{Department of Physics \& Astronomy, University of Manchester, Oxford Road, Manchester, M13 9PL, United Kingdom}
\affiliation{Henry Royce Institute for Advanced Materials, University of Manchester, Manchester, M13 9PL, United Kingdom}
\author{N. R. Wilson}
\affiliation{Department of Physics, University of Warwick, Coventry, CV4 7AL, United Kingdom}
\author{N. D. M. Hine}
\affiliation{Department of Physics, University of Warwick, Coventry, CV4 7AL, United Kingdom}

\begin{abstract}
   In van der Waals heterostructures, the relative alignment of bands between layers, and the resulting band hybridisation, are key factors in determining a range of electronic properties. This work examines these effects for heterostructures of transition metal dichalcogenides (TMDs) and hexagonal boron nitride (hBN), an ubiquitous combination given the role of hBN as an encapsulating material. By comparing results of density functional calculations with experimental angle-resolved photoemission spectroscopy (ARPES) results, we explore the hybridisation between the valence states of the TMD and hBN layers, and show that it introduces avoided crossings between the TMD and hBN bands, with umklapp processes opening `ghost' avoided crossings in individual bands. Comparison between DFT and ARPES spectra for the MoSe$_2$/hBN heterostructure shows that the valence bands of MoSe$_2$ and hBN are significantly further separated in energy in experiment as compared to DFT. We then show that a novel scissor operator can be applied to the hBN valence states in the DFT calculations, to correct the band alignment and enable quantitative comparison to ARPES, explaining avoided crossings and other features of band visibility in the ARPES spectra. 
\end{abstract}

\maketitle

\section{Introduction}
Ultrathin films of layered transition metal dichalcogenides (TMDs) form an important class of two-dimensional (2D) material. 
They have provided important results both in studies of fundamental physical behavior and in the development of optoelectronic and spintronic technological applications\cite{Splendiani2010,PhysRevLett.105.136805,Radisavljevic2011,Xu2014,Mak2016}.

Hexagonal boron nitride (hBN) frequently plays a crucial role in the fabrication and function of heterostructures and devices based on TMDs.
It is used as a dielectric barrier when gating a system, as a substrate, and as an encapsulator which is particularly important when a high-quality device is to be protected from atmospheric degradation\cite{Dean2010,Britnell2012,Courtade2018,PhysRevX.7.021026,Hua2021,Bandurin2016,Lee2022}. hBN is atomically flat, and has a lattice constant substantially different from those of the TMDs, thus avoiding disruption to the TMD crystal structure. Furthermore, it has a large band-gap and is generally expected to be electronically ``inert'' when used for encapsulation, with minimal impact on the electronic bandstructure of the TMD structure of interest.

In this work, we investigate the latter assumption above using electronic structure simulations. 
Comparison of band alignments between TMDs and graphene\cite{Wilson2017} to those between graphene and hBN\cite{Ogawa2019} suggest the possibility of the valence states of the TMDs and hBN could approach each other in other in energy. Therefore, could interaction between the hBN bands and the TMD bands affect the band structure and resulting electronic and optical properties of an hBN-encapsulated TMD? 
We focus on heterostructures where hBN is used to encapsulate the four most commonly studied semiconducting 2D TMDs, namely MoS$_2$, MoSe$_2$, WS$_2$ and WSe$_2$. 

Using linear-scaling density functional theory (LS-DFT), we are able to consider large-scale 2D material heterostructures with minimal strain. Furthermore, unfolding of energy bands back onto the primitive TMD and hBN unit cells allows direct comparison with spectral functions obtained experimentally using angle-resolved photoemission spectroscopy (ARPES) techniques.
The prediction of these LS-DFT calculations is that hybridisation between the bands of hBN and those of the TMD gives substantial avoided crossings in the unfolded spectral functions of the heterostructures, with `ghost' avoided crossings appearing in individual band spectra due to umklapp processes, similar to those previously observed in InSe/graphene heterostructures\cite{Graham2020}. 
Using standard semilocal DFT functionals, these avoided crossings are observed to appear close to the valence band edges of the TMDs, as the valence band maxima of hBN and the TMDs lie close in energy under the approximations of DFT.

However, band alignments between materials obtained from theory, even using advanced techniques, are often quite different from those found experimentally, such as using ARPES. 
They can change further upon creation of heterostructures, leading to very different conclusions about suitability for encapsulation.
To obtain a quantitative picture of the true band alignment from experiment, we analyze ARPES data for MoSe$_2$/hBN heterostructures. 
These show the band edges to be further separated in reality compared with the predictions of DFT. 
We therefore apply a scissor operator to the hBN valence states of the MoSe$_2$/hBN heterostructure to move them down in energy and correct the alignment of the MoSe$_2$ and hBN valence bands.
The resulting unfolded linear-scaling DFT spectra are then much more consistent with the ARPES results, with the relative positions of the MoSe$_2$ and hBN bands being in good agreement. We are then able to show that avoided crossings in the DFT bandstructure correspond to regions where the ARPES bands decrease significantly in visibility, suggesting a means to identify hybridised bands in heterostructures, even in the absense of a clearly-visible anticrossing in the ARPES results. 

This article is structured as follows: In Section \ref{sec:methods} we set out the first-principles methods and supercell details for the theoretical calculations of heterostructure properties in this work. In Section \ref{sec:DFT_results} the results of these calculations are presented, including charge-transfer induced potential drops across TMD/hBN heterostructures, and unfolded spectral functions showing avoided crossings in the band structure. In Section \ref{sec:muARPES} we show  ARPES data for heterostructures of MoSe$_2$ and hBN, and we discuss the application of a scissor operator to correct the band alignment. Finally, we apply the scissor-operator correction to other TMD/hBN pairings, informed by experimental band alignments of TMD/TMD heterostructures, to predict the band crossings that can be observed in such systems. We conclude our findings in Section \ref{sec:conclusions}. 

\section{Methods}
\label{sec:methods}
We utilise two complementary approaches to DFT: for smaller structures and simulation cell optimisation 
we use traditional plane-wave DFT calculations, whereas for large-scale heterostructures we utilise linear-scaling DFT \cite{Goedecker1999}.
\subsection{Finding interlayer distances using plane-wave DFT}
Since the properties of thin films of TMDs are sensitive to strain, the crystal structure parameters of the constituent TMD and hBN monolayers are taken from experiment\cite{Lynch1966,Bronsema1986,Schutte1987}. Other than applying small strains to the hBN layer to build commensurate supercells, these monolayer structures are kept fixed throughout our calculations. 

To determine the interlayer distance between the TMD and hBN layers, we used the plane-wave based VASP code\cite{VASP_1,VASP_2}, with standard PBE\cite{PBE} projector augmented wave method pseudopotential\cite{Blochl_1994}. We used an energy cutoff of 600 eV and sampled the Brillouin zone with a Monkhorst-Pack\cite{Monkhorst_1976} grid of $2\times2\times1$.  For accurate energetics of the interlayer interactions, we employ the optB88-vdW\cite{Klime_2009} non-local van der Waals functional for all calculations. This has been used in numerous previous studies of layered materials and their heterostructures and typically shows good agreement with experimental lattice constants and interlayer distances \cite{Constantinescu2015}.

Supercells for interlayer distance optimisation were generated using the code Supercell-core software\cite{Necio_2020}. Details of these supercells are shown in Table \ref{tab:interlayer_distances_supercells} in Appendix \ref{app:binding}. In Fig.~\ref{fig:binding_energies} in Appendix \ref{app:binding} we show the dependence of adhesion energy on chalcogen plane-hBN distance. A quadratic fit to the lowest energy points is then used to estimate the distance which gives the lowest energy, with the extracted optimal interlayer distances shown in Table~\ref{tab:supercells}.

\subsection{Spectral functions and potentials in heterostructures using LS-DFT}
Having fixed the monolayer structures and interlayer distances as set out above, we proceed to generate larger heterostructure supercells with very low strain and specific choices of relative twist. The resulting sizes necessitate calculations using the linear-scaling ONETEP code\cite{ONETEP}. The ONETEP code combines an underlying basis of periodic sinc (psinc) functions, equivalent to plane-waves \cite{Mostofi2003}, with a minimally-sized set of support functions, referred to as Non-orthogonal generalised Wannier Functions (NGWFs), which are optimised in-situ during a calculation\cite{Skylaris2002}. In terms of the NGWFs, the single-electron density matrix can be expressed as:
\begin{equation}
\rho(\mathbf{r},\mathbf{r}') = \sum_{\alpha\beta} \phi_\alpha(\mathbf{r}) K^{\alpha\beta} \phi_\beta(\mathbf{r}') \;,
\end{equation}
where $K^{\alpha\beta}$ is the density kernel, a generalisation of occupancy numbers to a nonorthogonal representation. The NGWFs are kept strictly local with a chosen spherical cutoff radius centered on their parent atom. The ground state Kohn-Sham energy can be expressed as the minimum of a functional of the NGWFs and the density kernel, expressed as $E[\{\phi_\alpha\},\{K^{\alpha\beta}\}]$. The NGWFs also provide a useful localised representation for the construction of projectors to uniquely identify the states associated with each layer of a heterobilayer. This is utilised both in spectral function unfolding and in the application of the scissor operator to shift states associated with one of the subsystems. 

For the ONETEP calculations in this work we use a psinc-grid energy cutoff of 800~eV, a cutoff radius of 12$a_0$ for the NGWFs, with 13 NGWFs on W and Mo atoms, 4 on S and Se atoms and 5 on B and N atoms. No truncation is applied to the density kernel. Repeated images of the slab are placed 60~\AA~apart, with a cutoff applied to the Coulomb interaction in the out-of-plane direction to suppress any interaction between the repeated slabs\cite{Rozzi2006,Hine2011}. The PBE approximation to the exchange-correlation functional is used \cite{PBE}, and PAW potentials from the GBRV dataset\cite{Garrity2014}. We calculate unfolded spectral functions\cite{Popescu2012} for the heterostructures following a methodology which adapts the approach to the NGWF representation, as described previously\cite{Constantinescu2015}.
 
\subsection{Scissor Corrections}
\label{sec:scissormethods}

To correct band alignments in light of ARPES results in Sec.~\ref{sec:bandalignments}, we find it necessary to apply a layer-dependent scissor operator to the Kohn-Sham states \cite{Levine1989,Gonze1997}. A scissor operator traditionally works on the basis of adjusting the bandgap by rigidly shifting the whole conduction manifold with respect to the valence manifold. This can be written as the a additional term $\hat{H}_{\mathrm{scissor}}$ added to the Kohn-Sham Hamiltonian in the (non self-consistent) bandstructure calculation:
\begin{equation}
    \hat{H}_{\mathrm{scissor}} = \sum_{i=1}^{N_{\mathrm{v}}} \sigma_{\mathrm{v}} | \psi_{i\mathbf{k}} \rangle \langle \psi_{i\mathbf{k}} | + \sum_{i=N_\mathrm{v}+1}^{\infty}\sigma_{\mathrm{c}} | \psi_{i\mathbf{k}} \rangle \langle \psi_{i\mathbf{k}} | \; ,
\end{equation}
where $N_{\mathrm{v}}$ is the number of occupied orbitals in the valence band, and $\sigma_{\mathrm{v}}$ and $\sigma_{\mathrm{c}}$ are shifts of the valence and conduction bands respectively, and $|\psi_{i\mathbf{k}}\rangle$ are the Kohn-Sham eigenstates, here constituting sets of projectors which act to select an energy range.

It is often the case for a heterostructure that the two subsystems (i.e. layers) have unphysical offsets in their DFT bandstructure when compared to experiment. This is well-known to be the direct result of limitations of the predictive power of approximate exchange-correlation functionals, particularly in representing energy gaps, and the different extent to which such errors affect different materials. In modelling heterogeneous systems, then, we may need to be able to make a different adjustment to each subsystem, or in the simplest case, to adjust the energy alignment of one layer with respect to another. In this case it is necessary to individually adjust the subspaces associated with each of two layers, 
\begin{equation}
    \hat{H}_{\mathrm{scissor}} = \sum_L \sigma_{\mathrm{v}}^L \hat{P}_{\mathrm{v}}^L +
    \sum_L \sigma_{\mathrm{c}}^L \hat{P}_{\mathrm{c}}^L \; ,
\end{equation}
where $L=1,2$ labels the layers, and $\hat{P}^L_{\mathrm{v}}$ and $\hat{P}^L_{\mathrm{c}}$ are projectors for the valence and conduction states respectively of layer $L$. In many forms of DFT it is not straightforward to separate out these states, since the real space volume in which they exist overlaps significantly in the region between the layers. Fortunately, as long as the degree of hybridisation is not extreme, then for a localised basis, suitable projectors may be constructed using the same basis functions from which we construct the Kohn-Sham eigenstates. In particular, for an LS-DFT calculation, we can use the set of support functions (NGWFs, in ONETEP) which are used in the construction of the density matrix. Because NGWFs are atom-centered, a layer index is automatically associated with each function corresponding to the layer of the atom.

The projectors are thus:
\begin{equation}
    \hat{P}_\mathrm{v}^L = |\phi_\alpha \rangle K_L^{\alpha\beta} \langle \phi_\beta |
\end{equation}
and
\begin{equation}
    \hat{P}_\mathrm{c}^L = |\phi_\alpha \rangle (S_L^{\alpha\beta} -K_L^{\alpha\beta}) \langle \phi_\beta |
\end{equation}
where $|\phi_\alpha\rangle$ are the non-orthogonal generalised Wannier functions, the support functions from which the density matrix is constructed in ONETEP, $K^{\alpha\beta}$ and $S^{\alpha\beta}$ are the valence density kernel and the inverse overlap matrix in the NGWF representation, and $K_L^{\alpha\beta}$ and $S_L^{\alpha\beta}$ are the components of those corresponding to NGWFs localised on the atoms of layer $L$. Summation over repeated Greek indices is implied. 

This approach works well in systems where the manifold associated with the two layers are suitably disjoint, so that states in one layer are not to any significant degree built out of orbitals from the other. 

\section{Density functional theory calculations of TMD-hBN heterostructures}
\label{sec:DFT_results}
\subsection{Supercell details}
\begin{table}
    \centering
        \caption{Twist angles $\theta$, number of atoms $N$, and maximum magnitude of components of strain in TMD+hBN supercells used for the potential and spectral function calculations in Section \ref{sec:DFT_results}.}
    \label{tab:supercells}
    \begin{tabular}{c|ccc}
    \hline\hline & $\theta$ & $N$ &Max $|$strain$|$\\
        \hline MoS$_2$/hBN & 31.6$^{\circ}$ &518& 0.3\% \\
        MoSe$_2$/hBN & 29.0$^{\circ}$ &587& 0.01\%\\
        WS$_2$/hBN & 31.6$^{\circ}$ &518& 0.08\%\\
        WSe$_2$/hBN & 31.7$^{\circ}$&753& 0.0002\%\\
        \hline
    \end{tabular}
\end{table}
In Table~\ref{tab:supercells} we show details of the supercells used in the LS-DFT calculations of heterostructures. Appropriate supercells are identified on the basis of minimising strain in layer 2 (the hBN layer in this case), with layer 1 remaining unstrained (the TMD). The misalignment angle is optimised by minimising a loss function characterising deviation from zero strain. The optimisation is performed within a certain angle range (near 30$^\circ$ and near 0$^\circ$, in this work) by finding the best layer 2 cell as a function of angle, having first identified a suitable range of possible layer 1 supercells meeting criteria for appropriate size and aspect ratio for the simulation.
The result is larger supercells than those used in the DFT-based interlayer distance optimization, chosen both to further minimise the strain and also because for spectral function unfolding the supercell lattice vectors must be greater in magnitude than twice the diameter of the NGWFs used in the calculation.
As long as the supercells used are sufficiently large, all possible local stacking orders are sampled within the supercell, so we can expect that the optimal interlayer distance will have little dependence on twist angle and particular choices of supercell lattice vectors\cite{Constantinescu2015}. 
\subsection{Interlayer charge transfer}
\begin{figure}
    \centering
    \includegraphics[width = 0.99\linewidth]{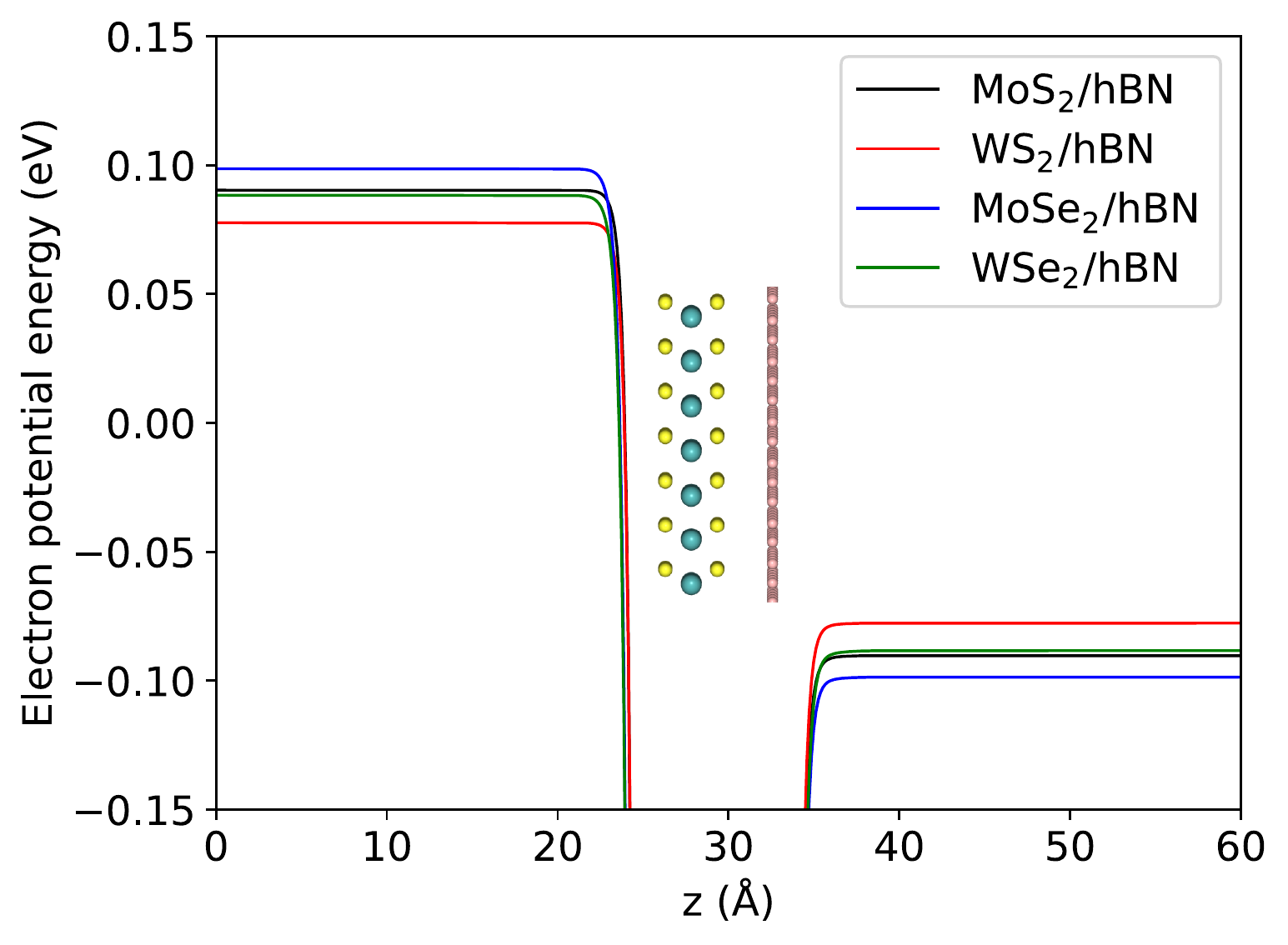}
    \caption{Electron potential energy (ionic and Hartree contributions) across TMD monolayer + hBN monolayer heterostructure (indicated schematically in inset), showing higher potential on TMD side due to net electron transfer from hBN to TMD.}
    \label{fig:charge_transfer}
\end{figure}
We first investigate the magnitude of interlayer charge transfer by examining the electrostatic potential in the heterostructure. In Fig.~\ref{fig:charge_transfer} we show the electrostatic electron potential energy calculated for the four TMD monolayer + hBN monolayer heterostructures. 
As demonstrated previously for layer asymmetric 2D material heterostructures\cite{Ferreira2021, Enaldiev2021, Weston2022, Wang2022}, hybridisation between conduction states of one layer (here TMD) with valence of the other (here hBN) can give rise to a net interlayer charge transfer between the two layers, with a potential drop across the heterostructure developing due to the resulting charge double-layer dipole. 

The result of this charge transfer is a shift of the band energies going from isolated monolayers to the heterostructure. In Fig.~\ref{fig:MoS2_hBN_combined} we show the DFT-calculated spectral functions of (left) isolated hBN and MoS$_2$ monolayers and (right) a monolayer hBN + monolayer MoS$_2$ heterostructure, with the bands unfolded by projection onto the unit cells of MoS$_2$ and hBN (yellow and blue, respectively). 
The effect of the aforementioned charge transfer can be clearly seen by comparing the valence band edges of each material, with the bands originating from MoS$_2$ moving up in energy and the hBN bands moving lower.
In the case of a TMD being sandwiched between layers of hBN, charge is transferred at both interfaces. This will give a symmetric potential profile. 

Due to the systematic underestimation of band gaps by DFT, we expect the energy difference between the unoccupied states of the TMD and the occupied hBN states to be larger than that predicted by DFT. Therefore, the potential energy differences $\sim$180~meV across the slab predicted by DFT should be understood as upper bounds for the likely size of the effect. The magnitude of overestimation can be understood by considering the energy differences as denominators in perturbation theory, reducing the amount of charge transferred by a given hybridisation strength. Comparisons between experiment and DFT calculations for a different system -- asymmetrically-stacked TMD homobilayers\cite{Weston2022, Ferreira2021, Wang2022} -- have shown that theory can still be used as a good estimate of the magnitude of charge transfer to be expected, in those cases the DFT overestimation is typically $\sim 10-20\%$. 
\subsection{Avoided crossings in unfolded spectral functions}
\begin{figure}
    \centering
    \includegraphics[width = 0.99\linewidth]{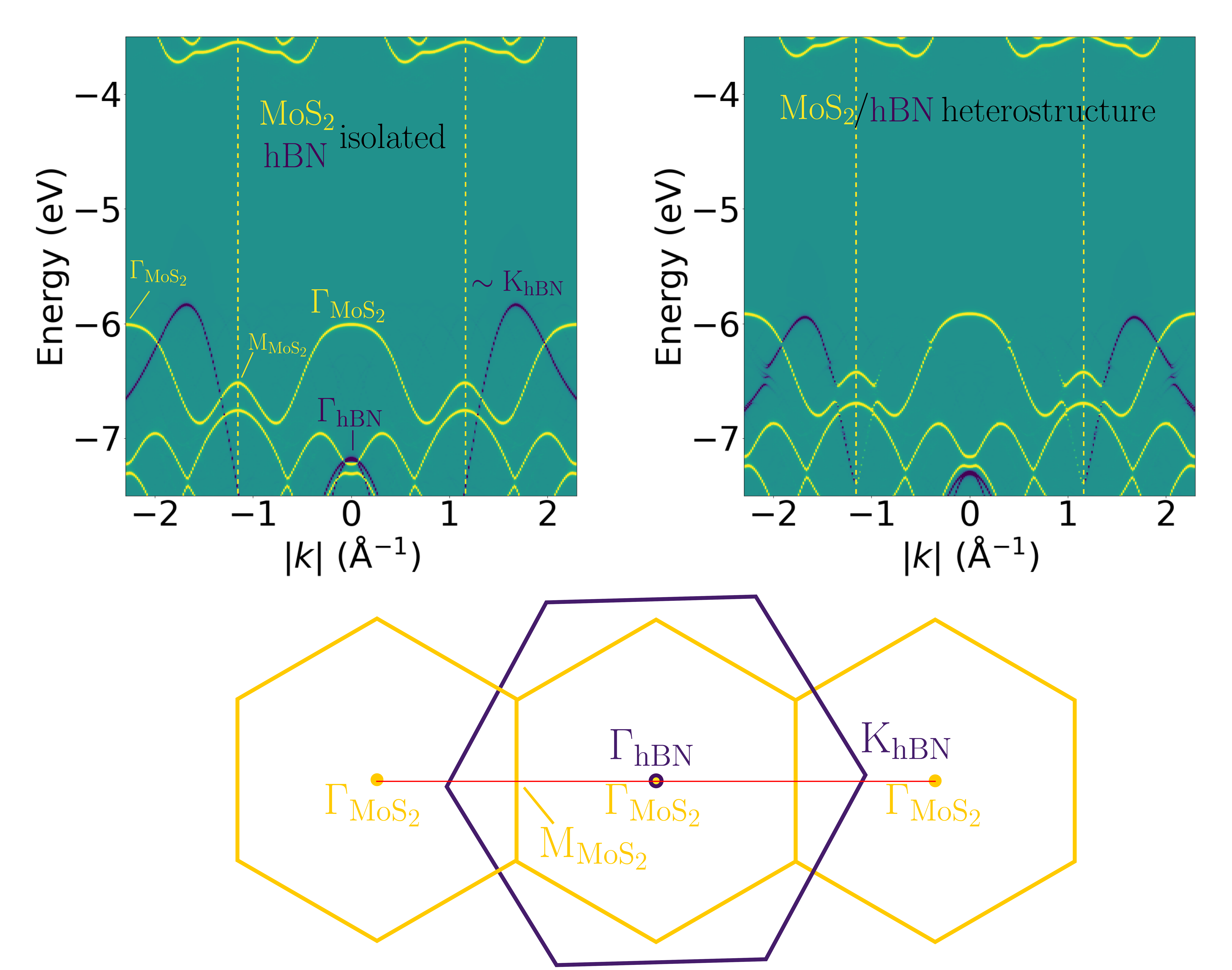}
    \caption{Upper panels: Unfolded spectral functions for isolated MoS$_2$ and hBN monolayers (left-hand side) and the two combined into a heterostructure (right-hand side). 
    Spectral functions are unfolded by projection onto the unit cell of monolayer MoS$_2$ (yellow) or hBN (blue). These are plotted along a k-point path, shown schematically below the panels, between $\Gamma$-point sites in the MoS$_2$ reciprocal lattice, which approximately passes through the K/K' points of hBN. 
    Yellow dashed lines indicate boundaries of first MoS$_2$ BZ.
    0~eV is set to the vacuum level for the isolated monolayers, and the mean of the vacuum levels on either side of the slab for the heterostructure.}
    \label{fig:MoS2_hBN_combined}
\end{figure}
The bands in Fig.~\ref{fig:MoS2_hBN_combined} are plotted along the k-point path for MoS$_2$ of $\Gamma$-M-$\Gamma^{'}$, where $\Gamma'$ is a point corresponding to one of the first `ring' of reciprocal lattice vectors, as shown in the inset to the figure. 
As the twist angle between the MoS$_2$ and hBN lattices is $\sim$30$^{\circ}$, this corresponds to a path in reciprocal space from $\Gamma$ to a point just beyond the K point of the hBN lattice.

Fig.~\ref{fig:MoS2_hBN_combined} demonstrates the existence of avoided crossings in the unfolded MoS$_2$ spectral functions appearing due to interlayer hybridisation between the MoS$_2$ bands and the hBN bands. This takes place near the valence band maximum of the hBN at the K-point, and thus outside the first Brillouin zone (BZ) of MoS$_2$. However, gaps in the spectra can still be seen inside the first MoS$_2$ BZ at momenta separated by a reciprocal lattice vector from the band crossings.
Such `ghost' avoided crossings, due to the similar distance in reciprocal space between neighbouring reciprocal lattice points of MoS$_2$ and between $\Gamma$ and K of hBN, have been observed in heterostructures of graphene and InSe\cite{Graham2020}. 
They are present in the systems described here despite the use of hBN as an inert encapsulation layer. However, for reasons explored further in Sec. \ref{sec:muARPES}, we do not expect them to have a problematic effect once the incorrect band alignment in DFT has been accounted for.
The avoided crossings in the MoS$_2$ bands are small near the valence band edge, and larger at deeper energies, since the orbital decomposition of the deeper bands has a stronger $p_z$-orbital contribution from the surface chalcogen atoms, promoting hybridisation with the hBN states\cite{Kormanyos2015}.
In the hBN bands hybridisation is minimal near the $\Gamma$ point where the wavefunction is dominantly of an in-plane $p_x,p_y$ character, and stronger towards the edge of the BZ where $p_z$ orbitals  dominate in the band-edge wavefunction\cite{Wickramaratne2018}.
In Fig.~\ref{fig:mos2_hBN_double} in Appendix~\ref{app:MoS2_hBN_double} we show the unfolded spectral functions for a heterostructure with hBN on both sides of MoS$_2$. 
The effect of the hBN on the projection of the bands onto the MoS$_2$ primitve cell is qualitatively similar to that found with just a single layer of hBN, but with larger avoided crossings due to the additional hybridisation arising from the hBN layer on the other side of the MoS$_2$.  
\subsection{Band edge valley shifts}
The orbital decompositions and symmetries of the wavefunctions in the various conduction and valence band valleys of the monolayer TMDs is such that stronger interlayer hybridisation occurs between states near the valence band $\Gamma$-point, and the conduction band Q-point, than at the K-point. 
This interlayer hybridisation pushes $\Gamma$-point hole states and Q-point electron states lower in energy with increasing number of layers in few-layer TMDs, and is a key contributor to the direct-to-indrect bandgap transition observed on going from the monolayer to thicker films in the TMDs\cite{PhysRevLett.105.136805, PhysRevB.98.035411}.

To understand whether interlayer hybridisation, here between a TMD monolayer and hBN, would affect the energy splitting between the TMD valleys, we show in Table~\ref{tab:valley_splittings} the differences in energies between the K- and $\Gamma$-point valence band valleys, and the Q- and K-point conduction band minima. 
These are shown for the isolated monolayers, as well as by unfolding the spectral functions of the TMD heterostructures by projection onto the TMD unit cells. In the valence band, the splitting between $\Gamma$ and K is reduced by the $\Gamma$-point states being pushed upwards in energy on hybridisation with the hBN bands, which lie lower in energy, while the K point is largely unaffected. 
hybridisation involving the conduction bands has a negligible effect on electron valley energies, since these bands are well-separated in energy from any hBN states.
\begin{table}
    \centering
        \caption{Energy differences (meV) between valence ($\Gamma$ and K) and conduction band (K and Q, the latter approximated as K/2) valleys for monolayer TMDs with and without monolayer hBN placed on one side. 
        Values in square parentheses for MoS$_2$ are for nearly-aligned MoS$_2$/hBN, while in rounded parenthesis for hBN placed on both sides.}
    \label{tab:valley_splittings}
    \begin{tabular}{c|cc|cc}
    \hline\hline
        & no hBN & with hBN & no hBN & with hBN \\
        & $E_{K}^{VB}-E_{\Gamma}^{VB}$&$E_{K}^{VB}-E_{\Gamma}^{VB}$&$E_{Q}^{CB}-E_{K}^{CB}$&$E_{Q}^{CB}-E_{K}^{CB}$ \\
        \hline MoS$_2$ & 210 & 185 [191](166) &70& 70 [64](71)\\
        MoSe$_2$ & 393 & 374& 121 & 125\\
        WS$_2$ &306 & 282 & 9& 15\\
        WSe$_2$ &541& 523 & 21& 15\\
        \hline
    \end{tabular}

\end{table}

\section{Accurate Band Alignments between hBN and TMDs}
\label{sec:bandalignments}

\subsection{ARPES results and comparison to DFT}
\label{sec:muARPES}

Given the challenges inherent in the theoretical modelling of 2D material heterostructures discussed above, we present in this section results from ARPES experiments on MoSe$_2$/hBN and WSe$_2$/hBN heterostructures, in order to extend our understanding of band alignments and resulting interlayer hybridisation at TMD/hBN interfaces. The samples were fabricated by standard mechanical exfoliation and polycarbonate-film-based dry transfer\cite{Zomer2014}, then annealed to 650 K for $\sim$4 hours in ultra-high vacuum before measurement. The resultant heterostructures have lateral dimensions of a few micrometres: to investigate these small samples, ARPES with sub-micrometre spatial resolution ($\mu$ARPES) measurements were made at the Spectromicroscopy beamline of the Elettra synchrotron. Linearly polarised light, with a photon energy of 27 eV and a 45\textdegree\ angle of incidence, was focused to a sub-micrometre diameter beam spot \cite{Dudin2010}.  During $\mu$ARPES experiments the sample temperature was 100$\;$K. 

\begin{figure*}
    \centering
    \includegraphics[width = 0.99\linewidth]{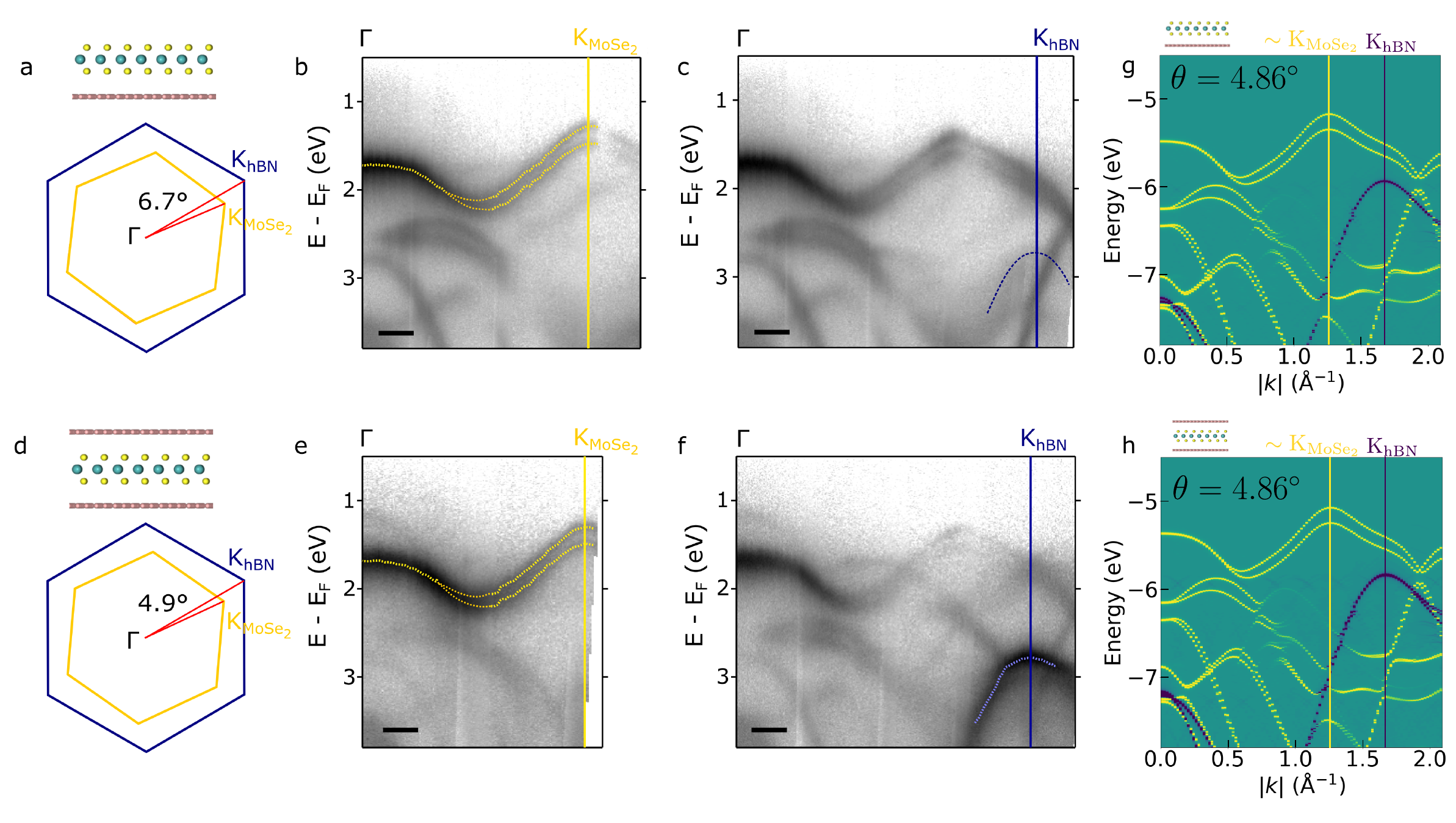}
    \caption{Comparison of $\mu$ARPES and unfolded DFT spectra (with no scissor correction applied at this stage) for MoSe$_2$/hBN heterostructures. (a) Schematics of MoSe$_2$/hBN heterostructure and of MoSe$_2$ and hBN Brillouin zones. (b) and (c) slices from $\mu$ARPES spectra along K$_{\mathrm{MoSe}_2}$ and K$_{\mathrm{hBN}}$ directions, respectively. (c)-(f) as (a)-(c), but for a region of the heterostructure with a hBN monolayer on top of the MoSe$_2$. These are shown with LS-DFT spectra along $\Gamma$-K$_{\mathrm{hBN}}$ for MoSe$_2$ nearly aligned with hBN, for monolayer hBN on one side (g) and both sides (h) of MoSe$_2$.}
    \label{fig:MoSe2_ARPES}
\end{figure*}
$\mu$ARPES spectra from a MoSe$_2$/hBN heterostructure are shown in Fig.~\ref{fig:MoSe2_ARPES}. At this photon energy, ARPES is sensitive to the top few atomic layers of the sample, allowing investigation of both the hBN and TMD valence band dispersions, and fitting the spectra enables direct extraction of the band parameters in each layer \cite{Wilson2017}. Spectra are shown from a monolayer of MoSe$_2$ lying at a twist angle of ${\sim}7^{\circ}$ on a multi-layer flake of hBN, as shown schematically in Fig.~\ref{fig:MoSe2_ARPES} (a) with spectra in (b) and (c), and of the same flakes with a monolayer of hBN on top, as shown schematically in Fig.~\ref{fig:MoSe2_ARPES} (d) with corresponding spectra in (e) and (f). The monolayer hBN is at a twist angle of ${\sim}5^{\circ}$ to the MoSe$_2$. In the $\mu$ARPES spectra, intensity can be seen corresponding to the valence bands in MoSe$_2$ and hBN, but there are also features which correspond to photo-emitted electrons that have been diffracted as they are emitted from the heterostructure\cite{Ulstrup2020}. For simplicity, our analysis concentrates on fitting the dispersion of the upper valence band in MoSe$_2$ (overlaid yellow lines) and hBN (blue lines) towards their respective BZ corners at $\bf{K}_{MoSe2}$ and $\bf{K}_{hBN}$ respectively. 

Since the positions of avoided crossings between bands in reciprocal space, and those of their `ghost' counterparts, depend on the twist angle between the layers\cite{Graham2020}, a comparison between experiment and theory requires simulations to use a supercell with a twist angle reasonably close to those found at the interfaces between the layers in the experimental device. We therefore construct a new supercell of a MoSe$_2$-hBN heterostructure. We look for a supercell with a twist angle between $4^{\circ}$ and $7^{\circ}$, finding a suitable candidate with a twist angle of 4.86$^{\circ}$. The resulting singly- and doubly-encapsulated models have 670 and 1028 atoms respectively, and a maximum component of the strain tensor of 0.9\%. LS-DFT calculations and spectral function unfolding were carried out on these models, and the resulting bandstructures are plotted alongside the experimental spectra in Fig.~\ref{fig:MoSe2_ARPES}.

In Table~\ref{tab:ARPES_DFT_alignments} we quantitatively compare valence band alignments extracted from experiment and theory. Whilst there are clear qualitative similarities between the DFT calculated spectral functions and the experimental spectra, in the DFT calculations the valence band edge of the hBN lies around 0.8~eV below that of the MoSe$_2$, compared to an energy difference of close to 1.5~eV in the $\mu$ARPES spectra. This difference highlights the well-known difficulty of quantitative determination of band alignments from first-principles calculations. This error, which we will proceed to correct by application of the scissor operator in the next section, precludes accurate identification of locations in the bandstructure where hybridisation between hBN and the TMD might be expected to be observed in ARPES.

\begin{table}
    \centering
        \caption{Comparison of $\mu$ARPES binding energies (B. E., eV) with DFT unfolded spectral function energies (eV) for (i) MoSe$_2$ on top of hBN (MoSe$_2$/hBN, bottom hBN is bulk for $\mu$ARPES, monolayer hBN for DFT) and (ii) as (i) but with an additional monolayer on top of the heterostructure (hBN/MoSe$_2$/hBN).}
    \label{tab:ARPES_DFT_alignments}
    \begin{tabular}{c|cc|cc}
    \hline\hline
    &\multicolumn{2}{c}{MoSe$_2$/hBN} & \multicolumn{2}{|c}{hBN/MoSe$_2$/hBN}\\
        & B. E. & $E_{\mathrm{DFT}}$ & B. E. & $E_{\mathrm{DFT}}$  \\
        \hline $\Gamma_{\mathrm{TMD}}$ & $1.71\pm0.03$& $-5.48$ & 1.68$\pm0.03$ &$-5.37$\\
        K$_{\mathrm{TMD},1}$ & $1.28\pm 0.03$&$-5.11$ & 1.30$\pm0.03$&$-5.01$\\
        K$_{\mathrm{TMD},2}$ & $1.48\pm 0.03$&$-5.29$ & 1.50$\pm0.03$&$-5.19$\\
        K$_{\mathrm{BN}}$ & $2.79\pm 0.05$ & $-5.94$& 2.78$\pm0.03$&$-5.84$\\
        \hline
    \end{tabular}

\end{table}

\subsection{Scissor correction}

Ideally, the incorrect valence band offset between MoSe$_2$ and hBN in the DFT results would be ameliorated by utilising a higher level of theory than the semilocal DFT employed here. However, for the large-scale models under investigation here, methods based on the GW approximation would be highly computationally expensive\cite{Deslippe2012}. A possible solution would be to use the offset of band edges relative to vacuum, obtained from GW calculations on monolayers to construct a correct offset, but we observe that this overestimates the required correction. Existing G$_0$W$_0$ results\cite{DTU_database_1,DTU_database_2} substantially overestimate the misalignment of the MoSe$_2$/hBN valence band edges, giving an energy difference $\sim$2.4~eV.

Motivated by the need to compare unfolded LS-DFT spectra to the $\mu$ARPES results with correct band alignments to investigate possible hybridisation, we apply a scissor operator to the LS-DFT Hamiltonian, keeping both the NGWFs and the density obtained before correction fixed. The requirement of separate manifolds of states associated with different layers in the scissor operator described in Sec. \ref{sec:scissormethods} is well-adhered-to by a 2D heterostructure. In the case of the MoSe$_2$ heterostructure discussed here, to adjust the relative position of the hBN valence bands to match the band alignment observed in $\mu$ARPES it is only necessary to have one nonzero component in the $\sigma$ values, namely $\sigma_\mathrm{v}^{\mathrm{hBN}}$. This is chosen to be negative to shift the hBN states to the appropriate position and is set as $\sigma_\mathrm{v}^{\mathrm{hBN}}=-0.65$~eV.
\begin{figure}
    \centering
    \includegraphics[width = 0.99\linewidth]{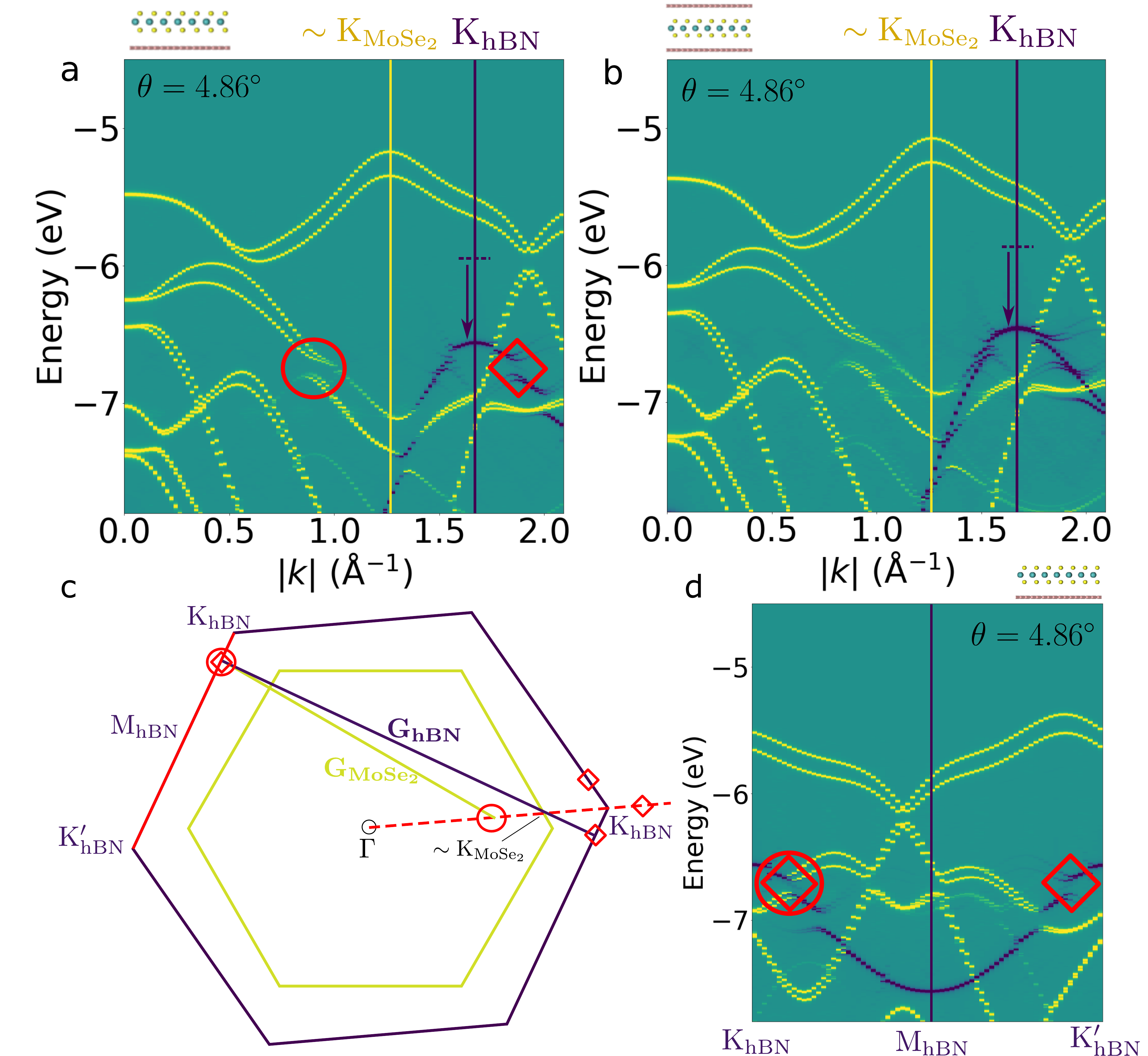}
    \caption{Scissor-corrected unfolded DFT spectra for MoSe$_2$/hBN heterostructures. Panel a: hBN on one side of MoSe$_2$, the circle marks the `ghost' avoided crossing seen in $\mu$ARPES data in Fig.~\ref{fig:MoSe2_ARPES}, the square its counterpart in the hBN projection. Panel b: hBN on both sides of MoSe$_2$. A scissor-operator is applied to move hBN valence states down in energy by 0.65~eV, as indicated schematically by inset arrows.  Panel c: Schematic of hBN and MoSe2 Brillouin zones (BZs). The red dashed line is the path along which unfolded DFT spectra are plotted in panels a and b, with the circle indicating the approximate position of the `ghost' avoided crossing in the MoSe$_2$ projection. The avoided crossing in the projection of the spectral function onto hBN (indicated by a square) lies along the K-M path of the hBN BZ, as well as being a reciprocal lattice vector of MoSe$_2$ away from the `ghost' crossing in the MoSe$_2$ projection inside the 1st MoSe$_2$ BZ. It is therefore possible to see the avoided crossing itself between the MoSe$_2$ and hBN bands by plotting along the path indicated by the solid red line between K and K$'$ of hBN, as shown in panel d. }
    \label{fig:MoSe2_sc}
\end{figure}
In panels a and b of Fig.~\ref{fig:MoSe2_sc} we show the resulting scissor-corrected bands, with the hBN valence states now lying in a region similar to where they are seen in the $\mu$ARPES results. 
The avoided crossing previously seen in the unfolded LS-DFT spectra is now located at approximately $-6.8$~eV for $|k|\sim 1.0$~\AA$^{-1}$.
Although the clear anticrossing behaviour exhibited by DFT is not clear to see in the $\mu$ARPES, the location of the anticrossing is evidently consistent with a strong decrease observed in the visibility of the $\mu$ARPES bands, which occurs roughly halfway between $\Gamma$ and $\mathrm{K_{hBN}}$ at binding energy $\sim$3~eV.
It may be that hybridisation of the TMD with a multi-layer flake of hBN rather than a monolayer, in the $\mu$ARPES results, obscures the anticrossing behaviour.
The origin of this `ghost' avoided crossing can be thought of as a point in somewhere in reciprocal space where the MoSe$_2$ and hBN bands coincide and hybridisation gives an avoided crossing.
This point can then be mapped to elsewhere in the BZ via umklapp processes. The exact location of this point cannot be inferred from a cut along $\Gamma$-K$_{\mathrm{hBN}}$ alone. 
In panel c of Fig.~\ref{fig:MoSe2_sc}, we identify where the `real' counterpart of the `ghost' avoided crossing may be found. 
The `ghost' crossing is found in the MoSe$_2$ projection of the DFT spectra within the 1st MoSe$_2$ BZ (marked by a circle). Beyond the K point of the hBN BZ an avoided crossing of a similar size and energy can be seen in the hBN-projected bands (marked by a square). 
These `ghost' crossings can be mapped, via umklapp processes using the reciprocal lattice vectors of the corresponding MoSe$_2$ and hBN primitive cells, to a point along the K-K' path on the edge of the 1st BZ of hBN. 
At this point, both MoSe$_2$ and hBN bands are present as the `real' counterpart of the `ghost' avoided crossing identified in ARPES and DFT, as shown in panel d of Fig.~\ref{fig:MoSe2_sc}. 

The positions of the `ghost' crossings are marked in panels a and b of Fig.~\ref{fig:MoSe2_sc} and their equivalent points in panel d are marked with the same symbols.
This illustrates the mapping of the avoided crossing from the edge of the 1st hBN BZ to a point within the 1st MoSe$_2$ BZ via umklapp processes. The relationship between these points is:
\begin{equation}
\underline{R} \, \mathbf{k}^{\mathrm{a.c.}}_{\mathrm{hBN}} + 
\mathbf{G}_{\mathrm{hBN}} =
\mathbf{k}^{\mathrm{a.c.}}_{\mathrm{MoSe2}} + 
\mathbf{G}_{\mathrm{MoSe2}},
\end{equation}
where $\mathbf{k}^{\mathrm{a.c.}}_{\mathrm{hBN}/\mathrm{MoSe_2}}$ and $\mathbf{G}_{\mathrm{hBN/MoSe_2}}$ are the k-space positions of the `ghost' avoided crossings and reciprocal lattice vectors of MoSe$_2$/hBN, respectively, and $\underline{R}$ rotates the position of the hBN avoided crossing by 120$^{\circ}$ about $\mathbf{K_{hBN}}$.

The hybridisation resulting in the avoided crossings discussed here is substantial, opening gaps $\sim 200$~meV. 
This arises due to the favourable orbital character of the bands involved at the crossings, with strong $p_z$ contributions from both MoSe$_2$ and hBN\cite{Kormanyos2015,Wickramaratne2018}.

\subsection{Scissor correction for other TMD/hBN pairs}
\begin{figure*}
    \centering
    \includegraphics[width = 0.99\linewidth]{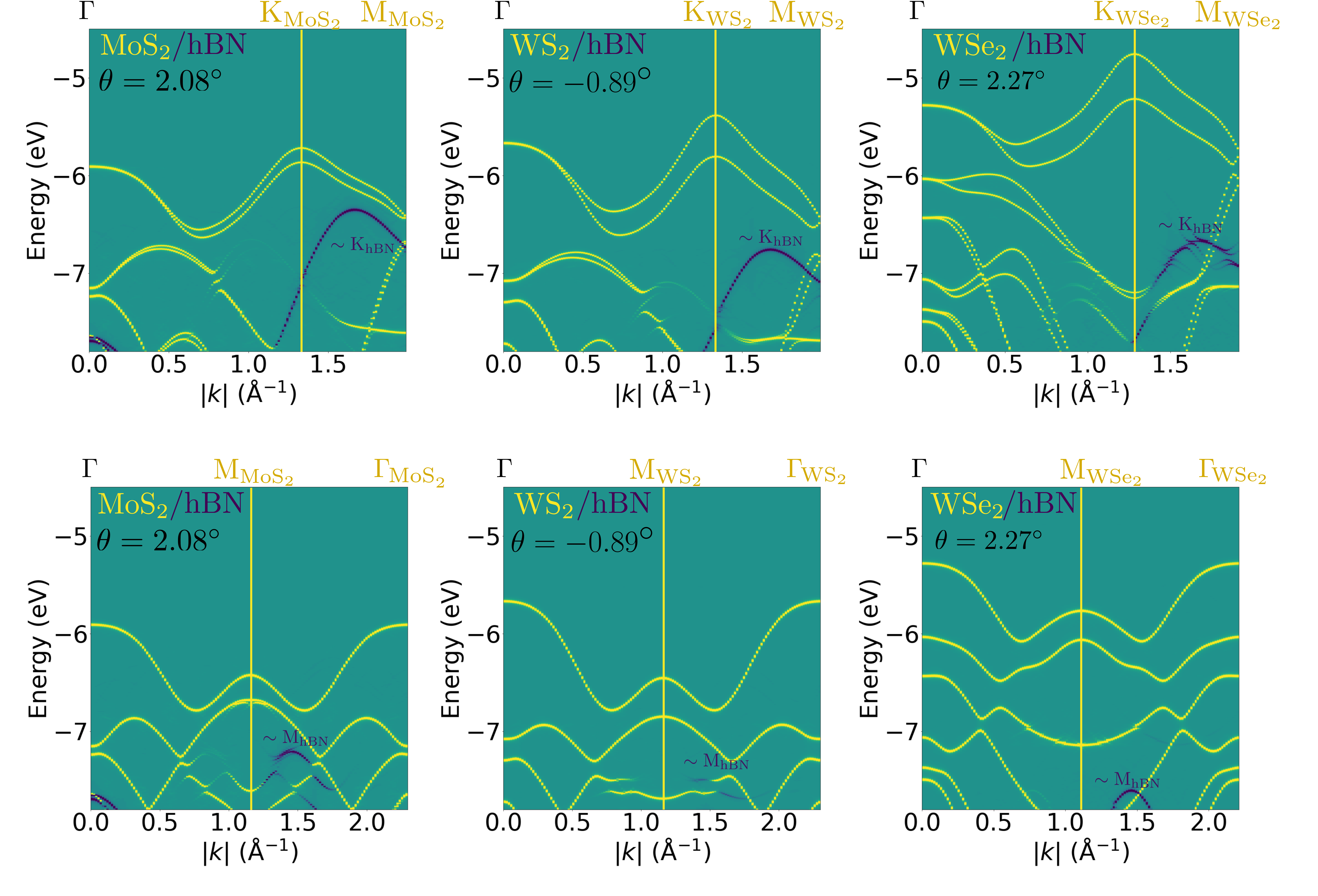}
    \caption{Unfolded DFT spectral functions for nearly-aligned TMD/hBN pairs with scissor correction applied to hBN, with the magnitude of the scissor operator determined approximately from experimentally found band alignments between the TMDs.}
    \label{fig:other_tmds}
\end{figure*}
\begin{table}
    \centering
        \caption{Twist angles $\theta$, number of atoms $N$, and maximum magnitude of components of strain in nearly-aligned TMD+hBN supercells used for spectral function calculations presented in Figs. 3-5.}
    \label{tab:supercells_0deg}
    \begin{tabular}{c|ccc}
    \hline\hline
        & $\theta$ & $N$ &Max $|$strain$|$\\
        \hline MoS$_2$/hBN &2.08$^{\circ}$ &1052& 0.4\% \\
        MoSe$_2$/hBN & 4.86$^{\circ}$ &670& 0.9\%\\
        WS$_2$/hBN & $-0.89^{\circ}$ &821& 0.02\%\\
        WSe$_2$/hBN & 2.27$^{\circ}$&855& 0.1\%\\
        \hline
    \end{tabular}

\end{table}
In Fig.~\ref{fig:other_tmds} we show unfolded DFT band spectra for nearly-aligned TMD/hBN pairs, where TMD=MoS$_2$,WS$_2$,WSe$_2$, to complement the MoSe$_2$/hBN spectra shown above.
We again apply the scissor operator to shift the hBN bands, leaving the TMD states unchanged. 
In choosing the magnitude of this shift, we seek to obtain a reasonable alignment between TMD and hBN valence bands for each pair. 
To find the energy differences between the valence band maxima of the other TMDs and that of hBN, we rely on experimental results giving energy differences between the valence band maxima of the TMDs\cite{Wilson2017,Hill2016,Stansbury2021}.
These are combined with the alignment between MoSe$_2$ and hBN extracted above to obtain an estimate of the valence band maximum separation for each TMD/hBN pair.
Comparison of these estimates with the band alignments for each pair found using DFT gives approximate values for an appropriate magnitude of scissor operator as $\sigma=-0.4$~eV for the  MoS$_2$/hBN pair, $\sigma=-0.85$~eV for WS$_2$/hBN, and $\sigma=-0.75$~eV for WSe$_2$/hBN. 

The behaviour seen in the MoSe$_2$/hBN pair is repeated in a similar manner in Fig.~\ref{fig:other_tmds} for the other TMD/hBN pairs. `Ghost' avoided crossings appear in spectral functions inside the 1st BZ of the TMD, due to hybridisation between the TMD and hBN bands occuring outside it. The predictions of the positions of these `ghost' crossings could be refined with a more precise knowledge of the TMD/hBN band alignment in question.

\section{Discussion and conclusions}
\label{sec:conclusions}
 In this work we have highlighted several features which must be taken into account in order to obtain an accurate knowledge of band alignment and the effects of hybridisation in 2D material heterostructures. 
 These include the need for a large simulation supercell, to minimise strain which would otherwise significantly modify bandstructures. The large sizes required are made feasible here through the use of linear-scaling DFT. 
 Charge transfer between the layers of the heterostructure will modify the band alignment of the materials, and this effect may be overestimated at DFT level, since the quasiparticle band gaps are underestimated. 
 Finally, an accurate description of the band alignments themselves is necessary. Estimates obtained from DFT will suffer from band-gap underestimation, while GW calculations will be extremely computationally expensive at the supercell sizes needed to minimise strain, and may in some cases overestimate the band misalignment. 
 
 From the experimental perspective it is possible to visualise the valence band spectra of heterostructures using ARPES, and we take advantage of this approach to correct our DFT band energies using a scissor operator with parameters estimated from the alignment of MoSe$_2$ bands with hBN in ARPES results. 
 Care must be taken in using ARPES for this, as the resolution of spectra is reduced away from band edges due to lifetime effects, and quasiparticle band energies can be affected by doping\cite{Nguyen_2019}.
 
 In conclusion, we have combined plane-wave and linear-scaling DFT with $\mu$ARPES experiments to examine and understand the band alignment and interlayer hybridisation in TMD/hBN heterostructures. We find that the energy differences between the valence bands of hBN and the TMDs allow for hybridisation between them, with avoided crossings appearing in unfolded DFT spectra as a result. `Ghost' avoided crossings appear in the spectra of the constituent layers away from the bands of the other material due to umklapp scattering where the point is a reciprocal lattice vector away from a band crossing\cite{Graham2020}.
 While DFT underestimates the misalignment between the TMD and hBN bands, we obtain a better estimate of the alignment using $\mu$ARPES for the MoSe$_2$/hBN pair, from which we can correct the alignment and positions of avoided crossings in the DFT by use of a scissor operator. 
 The scissor-corrected alignment allows changes in visibility of the $\mu$ARPES spectra to be interpreted as consistent with a `ghost' avoided crossing present in the DFT spectra.

The observation of these anticrossings in a 2D material system other than the InSe/graphene heterobilayer studied previously\cite{Graham2020} presents an opportunity to exploit their presence in a wider range of systems. Significant tunability in the position and size of the `ghost` avoided crossings is possible through the choice of material pairing and twist angle. By choosing favourable materials and tuning the band alignment, using an applied electric field or chemical doping\cite{Bao_2018}, it will be possible to position the avoided crossings such that they lie near the Fermi level. Signatures of novel behaviour could then be observed in transport experiments, such as through a Lifshitz transition whereby the topology of the Fermi surface changes as it passes through a `ghost' avoided crossing.

\acknowledgements{S.J.M. and N.D.M.H. acknowledge support from EPSRC grant EP/V000136/1. A.J.G. and N.Y. were supported by EPSRC studentships EP/R513374/1 and 1619532, respectively. F.F. and V.I.F. acknowledge EC-FET European Graphene Flagship Core3 Project, EC-FET Quantum Flagship Project 2D-SIPC, EPSRC grants EP/S030719/1 and EP/V007033/1, and the Lloyd Register Foundation Nanotechnology Grant. We thank Viktor Kandyba and Alessio Giampietri for assistance with experiments. Computing facilities were provided by the Scientific Computing Research Technology Platform of the University of Warwick through the use of the High Performance Computing (HPC) cluster Avon, the Computational Shared Facility of the University of Manchester, and the ARCHER2 UK National Supercomputing Service (https://www.archer2.ac.uk) through EPSRC Access to HPC project e672.}

\section*{Data availability} 
The data that support the findings of this study are openly available at the following URL: https://wrap.warwick.ac.uk/167461. 

\appendix
\begin{widetext}

\section{Binding energy curves}
\label{app:binding}
In Fig.~\ref{fig:binding_energies} we show DFT total energies as a function of interlayer distance (monolayer structures are kept fixed) found using plane wave DFT as described in Section \ref{sec:methods} in the main text. Details of the supercells used are set out in Table \ref{tab:interlayer_distances_supercells}. These results are used (via the dashed quadratic fits) to find the optimal interlayer distances for each TMD/hBN pair.
\begin{figure}
    \centering
    \includegraphics[width = 0.79\linewidth]{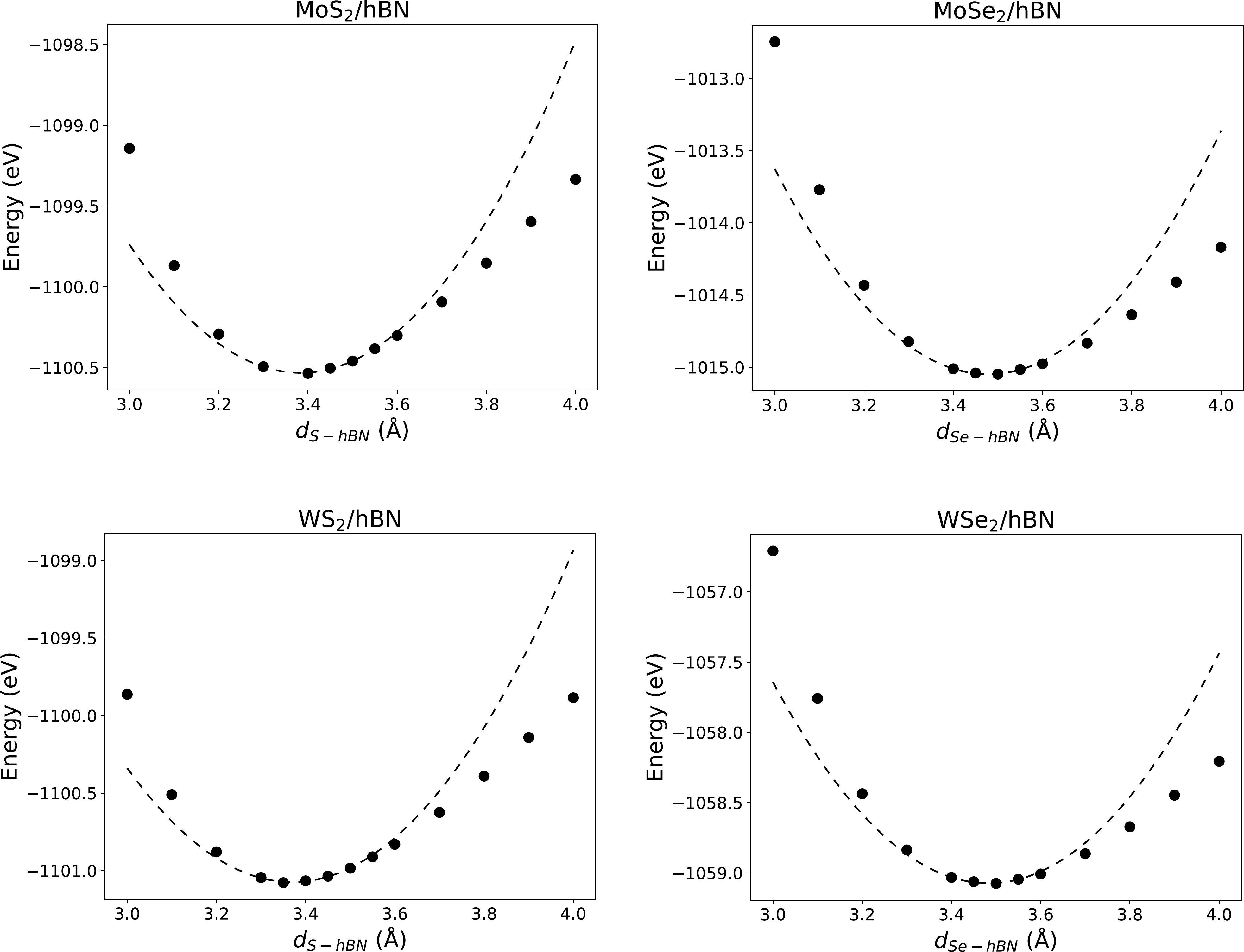}
    \caption{Dependence of TMD/hBN adhesion energies on chalcogen plane-hBN distance, for rigid monolayers.
    DFT data is represented by black dots where dashed lines are quadratic fits that were done around the minimum to extract the optimal chalcogen plane-hBN distance. }
    \label{fig:binding_energies}
\end{figure}
\begin{table}[ht]
    \centering
        \caption{Twist angles $\theta$, number of atoms $N$, maximum magnitude of components of strain in TMD+hBN supercells used for plane-wave DFT interlayer distance optimisation, and extracted interlayer distances (chalcogen to hBN-plane), $d$.}
    \label{tab:interlayer_distances_supercells}
    \begin{tabular}{c|cccc}
    \hline\hline
        & $\theta$ & $N$ &Max $|$strain$|$ & $d$ \\
        \hline MoS$_2$/hBN &21.6$^{\circ}$ &167& 0.2\% & 3.379~\AA \\
        MoSe$_2$/hBN &199.1$^{\circ}$ &161& 0.5\% &3.348~\AA\\
        WS$_2$/hBN &21.6$^{\circ}$ &167& 0.3\% & 3.361~\AA\\
        WSe$_2$/hBN & 199.1$^{\circ}$&161& 0.04\% &3.489~\AA\\
        \hline
    \end{tabular}

\end{table}

\section{Double-encapsulated MoS$_2$/hBN}
\label{app:MoS2_hBN_double}
In Fig.~\ref{fig:mos2_hBN_double} we show the counterpart of Fig.~\ref{fig:MoS2_hBN_combined} for a heterostructure with hBN layers on both sides of a MoS$_2$ monolayer. The bands associated with the two hBN layers result in two orthogonal linear combinations. For each crossing with MoS$_2$ bands, only one of these has a nonzero matrix element for hybridisation with MoS$_2$, and the other remains unchanged, bridging the gaps observed in the bandstructure for MoS$_2$ with single-sided encapsulation by hBN.  
\begin{figure}
    \centering
    \includegraphics[width = 0.79\linewidth]{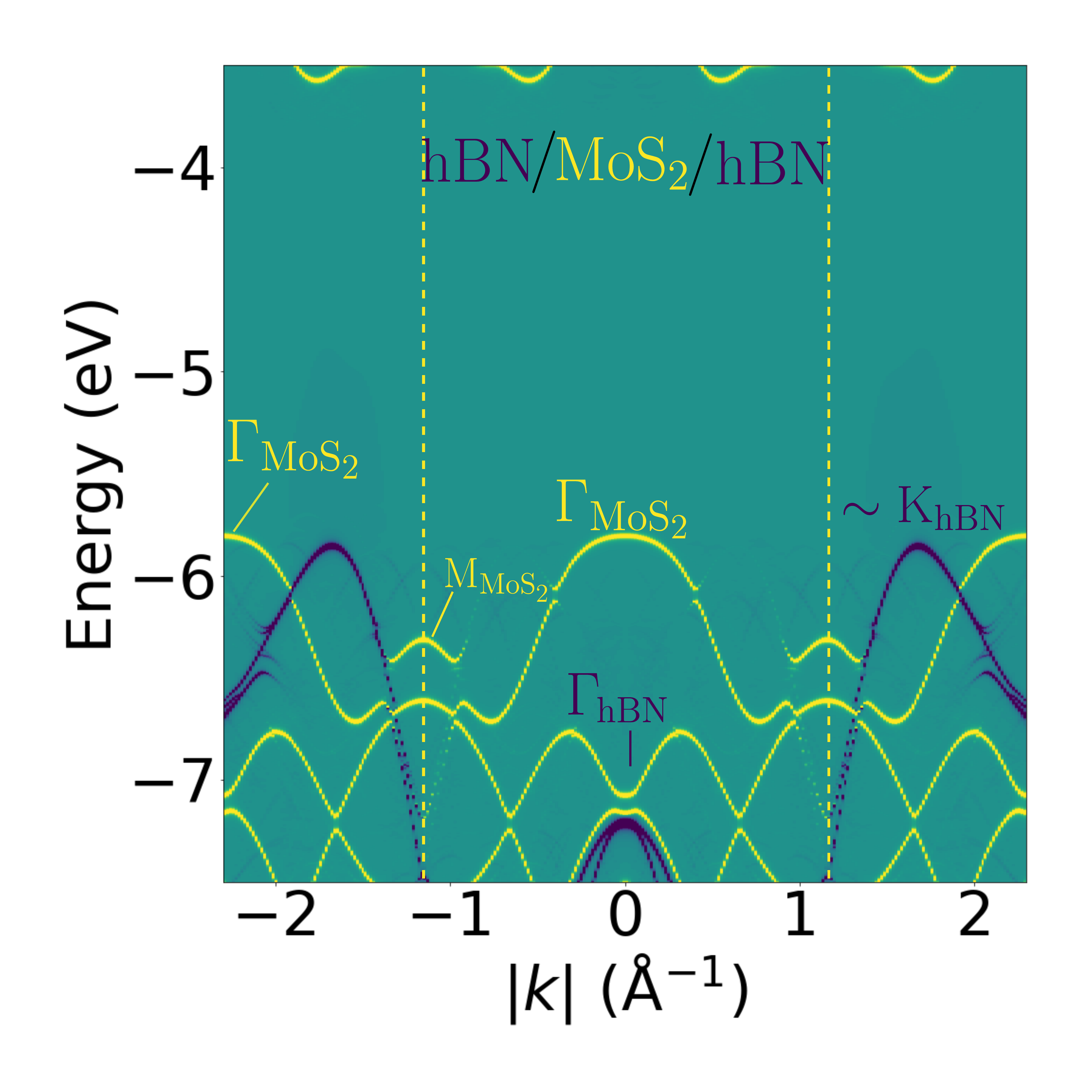}
    \caption{Counterpart of Fig.~\ref{fig:MoS2_hBN_combined} in the main text, but with hBN on both sides of the MoS$_2$.}
    \label{fig:mos2_hBN_double}
\end{figure}
\end{widetext}
%\bibliography{references.bib}

\end{document}